\documentclass[amsmath,amssymb,aps,prb,twocolumn,floatfix,superscriptaddress]{revtex4-2}

\usepackage{graphicx}
\usepackage[caption=false]{subfig}
\usepackage{standalone}
\usepackage{dcolumn}
\usepackage{bm}
\usepackage{xcolor}
\usepackage{physics}
\usepackage{amsthm}
\usepackage{enumitem}
\usepackage{mathtools}
\usepackage[colorlinks = true,linkcolor = red,citecolor = magenta]{hyperref}
\usepackage[sort&compress]{natbib}
\usepackage{orcidlink}

\DeclareGraphicsExtensions{.pdf,.eps,.png,.jpg,.mps}

\newcommand{\pcsadd}{Center for Theoretical Physics of Complex Systems, Institute for Basic Science (IBS), Daejeon, Korea, 34126}
\newcommand{\ustadd}{Basic Science Program, Korea University of Science and Technology (UST), Daejeon 34113, Republic of Korea}

\makeatletter
\renewcommand*{\fnum@figure}{{\normalfont\bfseries \figurename~\thefigure}}
\renewcommand*{\@caption@fignum@sep}{\textbf{:}}
\makeatother

\newcommand{\appropto}{\mathrel{\vcenter{
  \offinterlineskip\halign{\hfil$##$\cr
    \propto\cr\noalign{\kern2pt}\sim\cr\noalign{\kern-2pt}}}}}

\begin{document}

\title{Realization and characterization of an all-bands-flat electrical lattice}

\author{Noah Lape}
    \affiliation{Department of Physics and Astronomy, Dickinson College, Carlisle, Pennsylvania, 17013, USA}

\author{Simon Diubenkov}
    \affiliation{Department of Physics and Astronomy, Dickinson College, Carlisle, Pennsylvania, 17013, USA}

\author{L.Q. English}
    \email{englishl@dickinson.edu}
    \affiliation{Department of Physics and Astronomy, Dickinson College, Carlisle, Pennsylvania, 17013, USA}
    
\author{P.G. Kevrekidis}
    \email{kevrekid@umass.edu}
    \affiliation{Department of Mathematics and Statistics, University of Massachusetts, Amherst, Massachusetts 01003, USA}

\author{Alexei Andreanov\,\orcidlink{0000-0002-3033-0452}}
    \email{aalexei@ibs.re.kr}
    \affiliation{\pcsadd}
    \affiliation{\ustadd}

\author{Yeongjun Kim}
    \email{yeongjun.kim.04@gmail.com}
    \affiliation{\pcsadd}

\author{Sergej Flach}
    \email{sflach@ibs.re.kr}
    \affiliation{\pcsadd}
    \affiliation{\ustadd}

\date{\today}

\begin{abstract}
    We construct an electrical all-bands-flat (ABF) lattice and experimentally generate compact localized states (CLSs) therein.
    The lattice is a diamond (rhombic) chain and implemented as a network of capacitors and inductors, as well as voltage inverters (using operational amplifiers) in order to introduce a \(\pi\)-flux within each diamond.
    The network's normal modes split into three flat bands, and the corresponding CLSs can be excited in isolation via a two-node driving at the flat band frequencies.
    We also examine the role of the lattice edges and their interaction with the CLSs.   
    Finally, we compare the experimental results to tight-binding predictions and obtain very good agreement. 
    This analysis paves the way for further experimental implementations of ABF systems in electric networks, especially with an eye towards exploring their interplay with nonlinearity.
\end{abstract}

\maketitle

%----------------------------------------------------%

\section{Introduction} 
\label{introduction}

Flat band systems are tight-binding lattices in which one or more energy bands are dispersionless, e.g. perfectly flat~\cite{leykam2013flat,leykam2018artificial, danieli2024flat, rhim2021singular}.
In flat bands transport is completely suppressed even without disorder, leading to the compact localization (for finite-range hopping).
The macroscopically degenerate eigenstates of a flat band can take the form of conventional plane waves due to the translational symmetry, but they may also manifest as spatially compact, localized states known as compact localized states (CLS).
It is straightforward to obtain CLSs from transformations using combinations of plane wave eigenstates and vice versa.

CLSs have been sought in a variety of contexts including photonic lattices~\cite{nakata2012observation, mukherjee2015observation,kajiwara2016observation, vicencio2015observation, nguyen2018symmetry, ma2020direct},
polaritonic systems~\cite{baboux2016bosonic, masumoto2012exciton} or electrical chains~\cite{zhang2023non,wang2022observation,wang2019highly, zhou2023observation,chasemayoral2024compact}, among others.
As an important variation on the theme, the introduction of interactions~\cite{tovmasyan2013geometry, takayoshi2013phase, pelegri2020interaction, kuno2020interaction, nicolau2023many, kuno2020flat_qs, danieli2020many, vakulchyk2021heat, danieli2022many, tilleke2020nearest}, disorder~\cite{goda2006inverse,cadez2021metal, kim2023flat, lee2023critical, lee2023critical2}, or external magnetic fields~\cite{rhim2020quantum} in flat bands can induce nonperturbative,  nontrivial phenomena,
including the formation of compact nonlinear discrete states (such as breathers or standing waves)~\cite{vicencio2015observation,danieli2021nonlinear, danieli2018compact}, or the emergence of symmetry-breaking (or symmetry-restoring) bifurcations~\cite{vicencio2013discrete,nguyen2018symmetry}.

The absence of dispersion relies on fine-tuning of coupling interactions, so as to suitably
induce destructive interference.
To understand the underlying mechanisms that give rise to flat bands, multiple flat band generators for various lattice geometries and based on different construction principles have been proposed~\cite{maimaiti2017compact,hwang2021general, graf2021designing,ryu2024orthogonal,morfonios2021latent};
see, e.g., the detangling approach of Ref.~[\onlinecite{flach2014detangling}] or the constructive methodology of Ref.~[\onlinecite{morales2016simple}].
Such dispersionless features are particularly important in shaping the transport and overall dynamical properties of flat band lattices, justifying the particular interest in them~[\onlinecite{leykam2013flat},~\onlinecite{leykam2018artificial}].

Experimentally it is often challenging to observe the direct evidence of flat bands through the observation of CLSs, because this requires preparing (tuning), and observing the precise phase of the wave function in artificial lattice systems. Nevertheless, significant steps of recent progress have been reported in Ref.~\onlinecite{li2018realization, wang2022observation, mukherjee2015observation} 
Electrical circuits have become a particularly attractive testbed for artificial lattice engineering~\cite{lee2018topolectrical, sahin2025topolectrical}. 
With only a few passive elements ---capacitors, inductors, diodes--- one can build linear and nonlinear lattices that can support flat band linear spectra, as well as nonlinear solitonic excitations~\cite{english2012generation,english2010traveling, chen2018resonant}.
Adding active elements such as operational amplifiers (op-amps), one can moreover achieve tunable complex hopping phases, topological band structures, or PT symmetry~\cite{ezawa2019non}, nonlinearity~\cite{chasemayoral2024compact} and even interacting two-particles setups~\cite{zhou2023observation}. 
A scheme using only active elements has also been implemented~\cite{kotwal2021active}.
The versatility of electric circuits renders them a natural setting to realize flat bands and probing non-perturbative effects, like interactions, nonlinearities or disorder~\cite{chasemayoral2024compact,zhou2023observation}.

Creating an all-bands-flat (ABF) lattice is more challenging compared to a single flat band because of the higher degree of fine-tuning of the Hamiltonian, also possible requiring complex hoppings, and, consequently, higher sensitivity to perturbations. 
In an earlier realization of ABF using circuits~\cite{wang2022observation}, negative hoppings were realized by duplicating the network nodes, and interconnecting them, a common approach for synthetic gauge~\cite{lee2018topolectrical}.

In this manuscript we realize a particular ABF model in an electrical circuit.
More specifically, and crucially differing from our earlier work of~\cite{chasemayoral2024compact}, herein we examine a diamond chain with \(\pi\)-flux.
The fundamental underlying difference is that this constitutes a model which is known to be an all-bands-flat model (ABF), a particular extreme case of flat bands where all bands are flat~\cite{Kolovsky.108.L010201}.
Compared to systems with a single flat band, ABF lattices require more fine-tuning, and in particular hoppings with different signs resulting in modeling nonzero effective magnetic flux which penetrates individual plaquettes.
Such systems have only very recently started to be experimentally implemented, most notably in the photonic realm~\cite{kremer2020square, song2025observation, romancortres2025observation, yang2024realization}, and in electrical~\cite{wang2022observation} or superconducting~\cite{martinez2023interaction} circuits.
Hence further experimental platforms enabling their observation offer considerable additional insights, 
including the future promise towards an exploration of the interplay of ABF linear spectrum and nonlinearity.

We implement a different and simpler circuit than Ref.~[\onlinecite{wang2022observation}] to create a ``\(\pi\)-flux'', i.e. negative hoppings, using an inverting amplifier circuit with op-amps.
Our realization of ABF is validated through the observation of CLSs using a resonance via local sinusoidal driving targeted at the flat band frequencies. 
We observed a resonance response that is strongly localized on and near the driven site, and remains 
spatially compact throughout the lattice.
In addition to identifying the ABF CLS eigenmodes experimentally, we have developed a theoretical framework for modeling the corresponding lattice, and have observed very good qualitative and quantitative agreement with the experimental results.
This enhances our confidence that the present framework may prove to be a fertile one also for future studies.

Our presentation below is structured as follows.
In Section II we formulate the relevant experimental system and explain how the presence of the op-amps allows for the practical implementation of the ``negative coupling''.
The theoretical analogue of the model is introduced.
In Section III we present our experimental results on the frequencies and profiles of the CLS modes.
In Section IV we corroborate these findings on the basis of numerical simulations.
Finally, in Section V we summarize our findings and present our conclusions.

\section{Experimental System and Model}
\label{experimental}

\begin{figure}
    \includegraphics[width=\columnwidth]{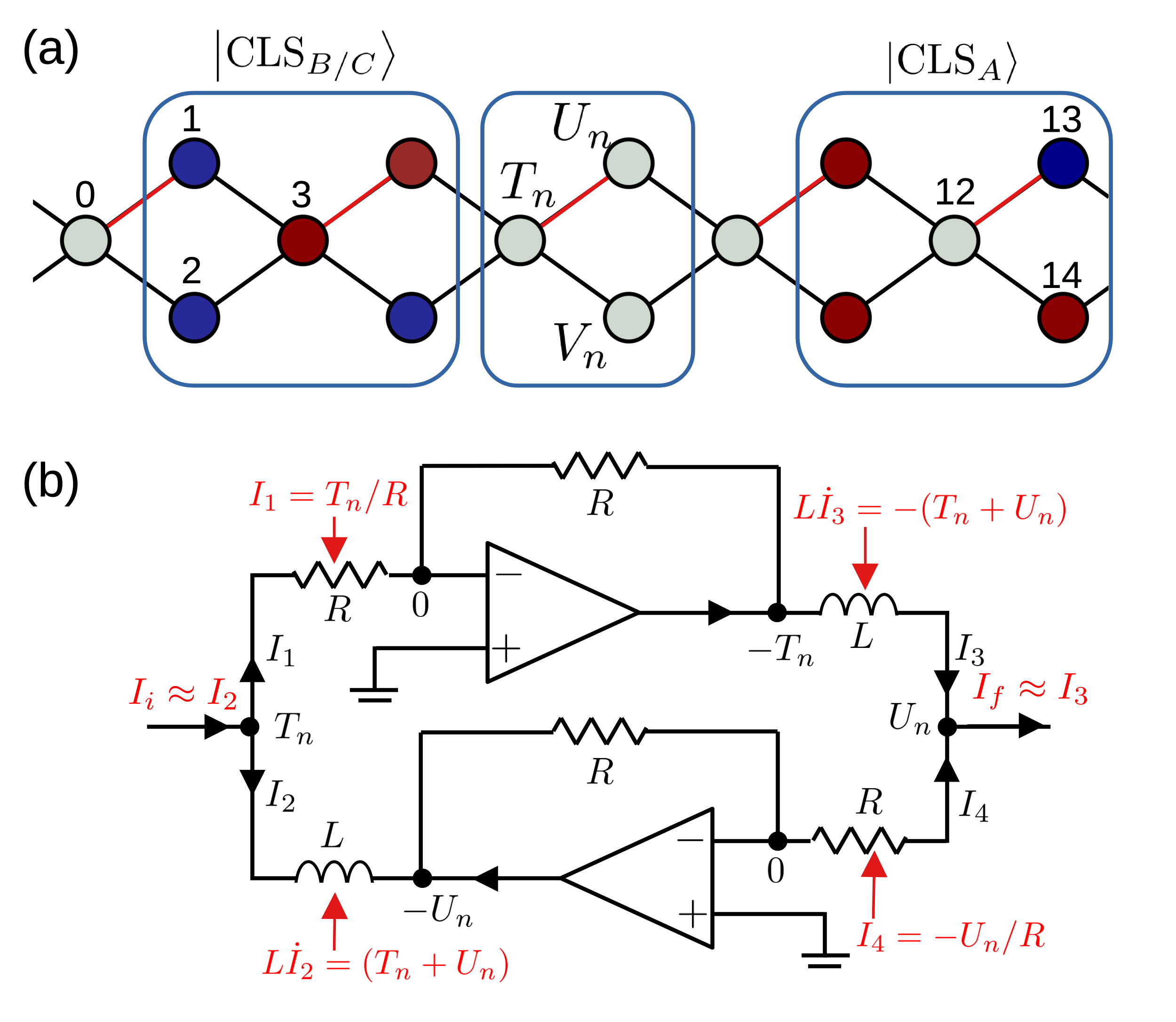}
    \caption{
        (a) Tight-binding representation of the basic diamond lattice, with its three-point basis, \(U_n, V_n, T_n\); also shown is our node numbering scheme.
        The CLS for the upper and lower flat band 
        (among the three FBs) is denoted as \(\ket{\mathrm{CLS}_{B/C}}\):
        The \(B/C\) CLSs have different amplitudes at the central site and peripheral sites (see main text).
        The CLS for the middle flat band \(\ket{\mathrm{CLS}_{A}}\) is characterized by a vanishing amplitude at the center site (site 12).
        In the context of the electrical circuit, circles represent \(LC\)-resonators, black lines depict inductors that couple two resonators, and red lines represent negative inductive coupling. 
        (b) Negative coupling is achieved through voltage inverters, here using op-amps.
        In (a), we see that the connection between \(T_{n}\) and \(U_n\) is negative - the red line is a placeholder for this circuit, explained in more detail in the text.
    }
    \label{fig:setup}
\end{figure}

We study a diamond chain with three sites (voltages) per unit cell 
encoded by site voltages, 
(\(T_n, U_n, V_n\)),
as shown in Fig.~\ref{fig:setup}(a) where one hopping per plaquette carries a phase \(\phi\) ensuring a \(\phi\)-flux threading each plaquette.
Here, black and red edges represent hoppings \(t\) and \(te^{i\phi}\), respectively. 
Its \(3\times3\) Bloch Hamiltonian in the \((T_k, U_k, V_k)\) basis reads: 
\begin{align}
    H(k,\phi)=
    t\begin{pmatrix}
        0 & 1 + e^{-i(k+\phi)} & 1+e^{-ik}\\
        1+e^{i(k+\phi)} & 0 & 0\\
        1+e^{ik} & 0 & 0
    \end{pmatrix}.
\end{align}
Setting \(\phi=\pi\) makes the red edges real negative \(te^{i\phi} = -t\).
Thus, if a single diamond is traversed in a clockwise loop, a phase of \(\pi\) accumulates - a hallmark of an Aharonov-Bohm cage~\cite{vidal1998aharonov, vidal2000interaction, douccot2002pairing}.
By direct diagonalization, it can be shown that at \(\pi\)-flux, all eigenvalues are \(k\)-independent, producing an all-bands-flat (ABF) spectrum \(E =  0, \pm t\) with CLSs spanning two unit cells as shown in Fig.~\ref{fig:setup}(a)~\cite{vidal1998aharonov}.

In the electrical circuit network analogue of the tight-binding model, the sites (nodes) represent \(LC\)-resonators, and hoppings (edges) represent inductors that couple two resonators in Fig.~\ref{fig:setup}(a).
Here, red lines now represent negative inductive couplings.
For the experimental implementation of the ABF lattice, we have modified an electrical diamond lattice described in Ref.~\onlinecite{chasemayoral2024compact}.
The key modification involves periodically introducing \(\pi\)-phase shifts by electrically implementing negative coupling edges.

The design of such a negative edge is shown in Fig.~\ref{fig:setup}(b), where the two nodes \(T_n\) and \(U_n\) are to be negatively coupled.
If these two nodes had been connected by a single (ordinary) inductor, the current \(I\) flowing from the left to the right node would be governed by \(L\dot I = (T_n-U_n)\).
We would like to modify this equation to read \(L\dot I_f = (-T_n-U_n)\), where \(I_f\) denotes the total current flowing into node \(U_n\).
This is accomplished by the top branch of the parallel circuit in Fig.~\ref{fig:setup}(b).
The voltage at the two ends of the inductor (top, right) has values of \(-T_n\) and \(U_n\), respectively, due to the inverting amplifier of unity gain.
The resistors connecting the inverting input node(-) of the operational amplifier (Op-Amp), and the one at the feedback loop which connects the output inverting node, are of equal value \(R\).
As we will argue shortly, the resistance \(R\) must also be sufficiently large.
The negative coupling needs to be bi-directional.
In other words, the current \(I_i\) flowing out of \(T_n\) node obeys \(L\dot I_i= (T_n+U_n)\).
This is accomplished by the lower branch of the parallel circuit.

The two currents \(I_i\) and \(I_f\) do not interfere with one another, because the input impedance of this inverter is given by the resistance \(R\) of the resistors, and this value is chosen to be much larger than the impedance of the inductor \(R \gg |i\omega L|\), where \(\omega\) is the driving frequency.
When this condition is met, very little current originating, for example, from the output of the lower Op-Amp is diverted to the upper branch and flows (almost) entirely into node \(T_n\), i.e. \(I_i \approx I_2\) in Fig.~\ref{fig:setup}(b).
Furthermore, for practical reasons, additional resistors \(R_G\) are used to connect every non-inverting op-amp input to ground (not shown in the diagram)
~\footnote{This resistance does not affect the op-amp if it is ideal. 
In practice, the impedance difference at two inputs creates small input bias voltage between two inputs: \(V_{B} = I_B/(R_{-} - R_G)\), distorting the output voltage.
To mitigate this effect, \(R_G\) is attached at the (+) input,  that \(R_G = R_-\), i.e. \(R_G = R/2\).
See Ref.~\onlinecite{TI_AN20_2013} for technical detail. Furthermore, note that \(R\) should not be too large, because then, parasitic capacitance at the inverting input induces unwanted \(\pi/2\) phase-shift and large noise at high frequency.}. 

The sites in Fig.~\ref{fig:setup}(a) represent \(LC\)-resonators, i.e., a parallel combination of an inductor and a capacitor connecting the node to ground.
We can choose their inductance to be infinitely large (open circuit), as justified below in Eq.~\eqref{eq:ham-abf-neat-form}, so that each \(LC\)-site reduces to a single capacitor \(C\).
Thus, the lattice is characterized by only one capacitance value \(C\) of the nodes to ground, and one inductance value, \(L\), of the coupling between nodes.
Applying Kirchhoff's voltage law at each node, we arrive at a system of linear differential equations of the form~\cite{chasemayoral2024compact},
\begin{align}
    \notag
    \ddot{T}_n+(4\omega_0^2) T_n &= \omega_0^2 (U_n-U_{n+1}+V_n+V_{n+1}), \\
    \label{eq:ham-abf}
    \ddot{U}_n+(2\omega_0^2) U_n &= \omega_0^2 (T_n-T_{n-1}), \\
    \notag
    \ddot{V}_n+(2\omega_0^2) V_n &= \omega_0^2 (T_n+T_{n-1}),
\end{align}
where \(\omega_0^2=1/(LC)\). 

Equations (\ref{eq:ham-abf}) describe the dynamics of the ABF lattice.
Using the Bloch-wave Ansatz, \(U_n = U(k) \exp[i(\omega t - kn)]\) (and similarly for \(V_n\) and \(T_n\)) we obtain the eigenvalue equation for the ABF Hamiltonian in reciprocal Bloch space:
\begin{align}
    \label{eq:ham-abf-neat-form}
    -\dfrac{\omega^2}{\omega_0^2}\vec\psi(k) = [H(k, \pi) - D]\vec\psi(k).
\end{align}
Here, the diagonal terms (onsite potentials) \(D = \mathrm{diag}(4, 2, 2)\) reflect the number of inductors coupled to the sites \(T_n, U_n\) and \(V_n\) and ensure \(\omega^2 > 0\).
Note that unlike the conventional chiral diamond chain~\cite{leykam2018artificial,morales2016simple}, unequal diagonal terms are present, yet the spectrum remains dispersionless.
While the chiral symmetry is formally broken due to the presence of nonzero diagonal terms, a frequency gauge can remove the diagonal terms for the majority sublattice \(U,V\) and therefore the middle flat band is still protected by 
partial chiral symmetry~\cite{chasemayoral2024compact,calugaru2022general}.

Note that a finite onsite inductance \(L_{0}\) at a site of the \(LC\)-resonator results in a shift of the resonance frequency of that site by \(\omega_0^2(L/L_{0})\), which appears as an additional onsite potential term in Eq.~\eqref{eq:ham-abf}.
If needed, one might also add site-dependent onsite inductances to tune the onsite potential of the model at each site.
This may be useful for future studies exploring the interplay of disorder through ``impurity nodes'' on the ABF lattice, 
generalizing earlier such studies in regular electrical
lattices; see, e.g.,~\cite{molina}.
in electrical lattices and of the interplay of two different types of localization (FB- and Anderson-induced)~\cite{dresselhaus2024tale}.
In our case, this tuning is not needed, since Eq.~\eqref{eq:ham-abf-neat-form} already describes ABF, without additional onsite potential tuning.

Solving the eigenvalue problem yields three eigenvalues:
\begin{align}
    \omega_A =\sqrt{2} \, \omega_0,\qquad
    \omega_{B/C} = \left(\sqrt{3\pm \sqrt{5}}\right) \omega_0.
\end{align}
Note the absence of \(k\) in these expressions, and hence the existence of three flat bands.

The CLSs of each flat band are shown in Fig.~\ref{fig:setup} (a).
Every CLS occupies two unit cells, forming an ``X'' shape with 4 peripheral sites \(U_{n}, V_{n}, U_{n+1}, V_{n+1}\) and 
a central one located between them at the \(T_n\)-site. 
For the middle flat band, the amplitude at the center site is zero, due to the partial chiral symmetry associated with the flat band.
On the other hand, for CLSs of the lower and upper flat bands, the central site and the peripheral sites are simultaneously nonzero.
The ratio of the amplitudes at the central site vs. the left peripheral sites (sites 1, 2 in Fig~\ref{fig:setup}(a)) are \((\sqrt5 \pm 1)\) for \(\omega_B, \omega_C\), respectively. 
For \(\omega_B\), the CLS (corresponding to the lower flat bands), \(\{U_n,V_n\}\) and \(U_{n+1}\) 
are in-phase with \(T_n\) site,
while for \(\omega_C\) (corresponding to the upper flat bands), they are in anti-phase with respect to \(T_n\).
In all cases, the voltages \(\{U_n,V_n\}\) of the left pair are in-phase with each other, whereas those of the right pair, \(\{U_{n+1},V_{n+1}\}\), are in anti-phase with each other.

\section{Experimental Results}
\label{sec:exp_res}

In our setup, we use \(L = 470\)~mH and \(C = 4.7\) nF.
The resistances of the Op-Amps are chosen as \(R = 10~\textrm{k}\Omega\), ensuring \(\omega L \ll R\) for the frequencies \(\omega\), which is an important criterion as described before (note that \(R_G = R/2 =  5~\textrm{k}\Omega\)). We use LF411 op-amps.
This results in the following predictions for three flat band frequencies: \(94\) kHz, \(152\) kHz, and \(245\) kHz.

To probe the CLS, the lattice is driven by a signal generator at one or two sites via small coupling capacitors (\(C_d = 3.5~\textrm{pF} \ll C)\)~\cite{chasemayoral2024compact}.
Driving at two sites enables the isolation of a single CLS in this system. 
Spectra are recorded by sweeping the drive frequency and measuring the response at a nearby site, while spatial profiles are captured simultaneously across all nodes using fast DAQ boards (NI 6133 cards, 2.5 MHz sampling).
The detailed spatio-temporal resolution of the relevant system is one of the advantages of the present setting.
As was shown in Ref.~[\onlinecite{chasemayoral2024compact}] via a transfer function analysis, once the drive frequency coincides with a flat band eigenvalue, the response of the entire network localizes to the CLS that involves the driven sites.
It is easy to see from the structure of the CLS, Fig.~\ref{fig:setup}(a), that driving two sites \(U_n\) and \(V_n\), with in-phase and out-of-phase, isolates a single CLS:
in-phase driving at \(U_n\) and \(V_n\) excites the CLS over cells from \(n\) to \((n+1)\), whereas anti-phase driving addresses the CLS over cells from \((n-1)\) to \(n\).
In this way, even though we are only exciting a single pair of $(U_n,V_n)$, we uniquely select a CLS mode pertaining to the suitable ABF associated with the drive frequency.

\subsection{Excitations in the bulk}
\label{sec:bulk}

\begin{figure*}[htpb!]
    \includegraphics[width=\textwidth]{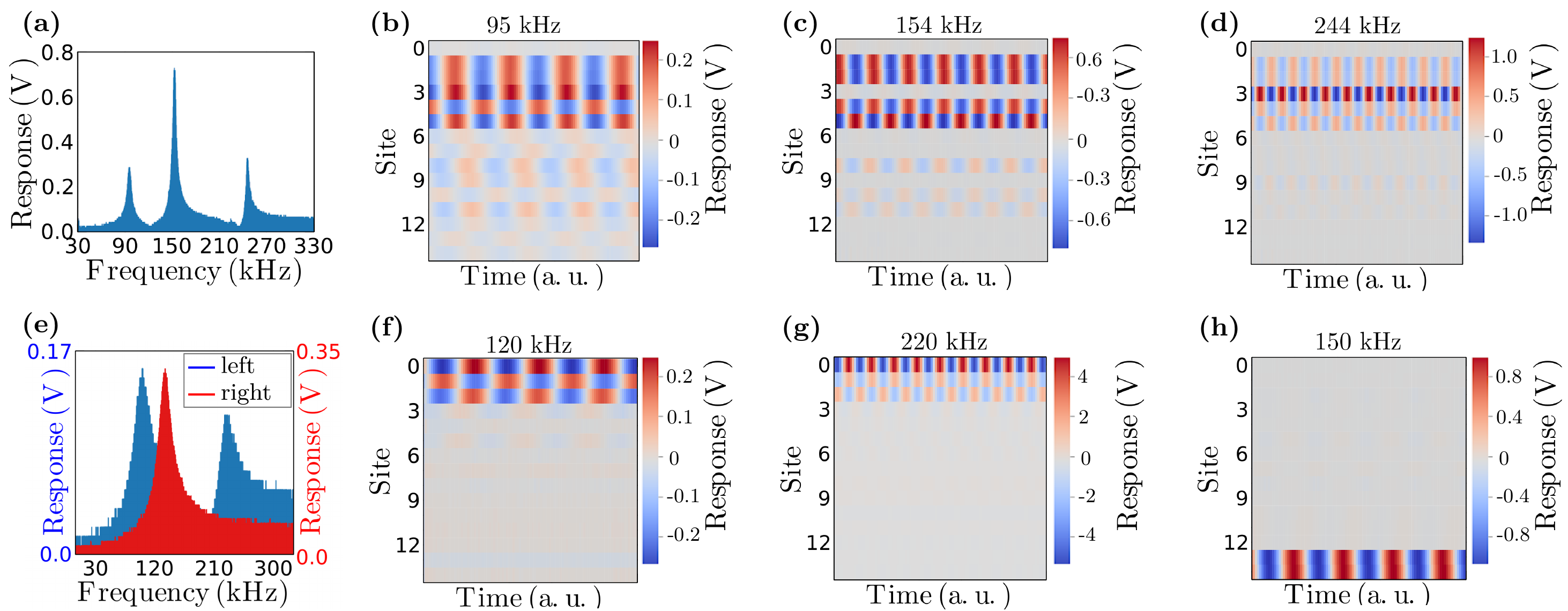}
    \caption{
        (a)-(d) The ring lattice with periodic boundary conditions is driven at the left-edge nodes 1 and 2 (\(U_0,V_0\)) in-phase and observed at \(V_0\).
        (See Fig.~\ref{fig:setup}(a) for site labeling).
        (a) The response at \(V_0\) vs. frequency.
        The spectrum shows three prominent peaks at 95~kHz, 154~kHz, and 244~kHz.
        (b)–(d) The full spatio-temporal signature of the CLS in steady state under sinusoidal driving at \(f = 95\)~kHz, 154~kHz, and 244~kHz, respectively.
        The time window (x-axis) ranges from 20~\(\mu\)s to 50~\(\mu\)s, capturing 3–10 oscillations \((3/f–10/f)\) in steady state.
        (e) The blue shade shows the frequency response when driving the left-edge nodes 1 and 2 (\(U_0\) and \(V_0\)) anti-phase and observing at node 2.
        Two peaks appear at 120~kHz and 220~kHz, in good agreement with the edge state eigenvalues calculation.
        The red shade shows the response when driving the right-edge nodes 13 and 14 (\(U_4\) and \(V_4\)) in-phase.
        (f)–(h) The full spatio-temporal signature of the CLS in steady state under sinusoidal driving at the left edge, for \(f = 120\)~kHz, 220~kHz, and at the right edge, 244~kHz, respectively.
        }
    \label{fig:Fig2}
\end{figure*}

We first study a ring (periodic boundary conditions) of five unit cells by connecting nodes 13 and 14 to node 0 in Fig.~\ref{fig:setup}(a).
We then inject a sinusoidal signal from a signal generator in sweep mode into one or two nodes of the lattice via a small coupling capacitor (\(C_d = 35\,\)pF).
The drive amplitude was varied between \(1\) V and \(10\) V, which lies well within the components’ specified operating ranges, ensuring a linear response throughout the experiment.

Figure~\ref{fig:Fig2}(a) shows the resulting spectrum from two-node driving (the pair of nodes, \(U_n, V_n\), are driven in-phase via two different coupling capacitors).
Three clear resonance peaks can be seen at \(95\) kHz, \(154\) kHz, and \(244\) kHz, which match the above theoretical prediction for flat band frequencies very well.
Single-node driving yields a very similar result in this configuration.
The broadening of the peaks is primarily due to the \emph{effective series resistance} \(R_{\mathrm{eff}}\approx 20~\Omega\) from the inductors, as analyzed previously in Ref.~[\onlinecite{chasemayoral2024compact}].

Now that the flat band frequencies have been experimentally ascertained and verified, we can use the signal generator at a fixed frequency to probe the lattice at those three frequencies sequentially.
Here we would like to capture the full spatio-temporal lattice dynamics by recording the voltage responses at each node, using the fast DAQ-boards.
We start with the lowest flat band frequency of 95 kHz, shown in Fig.~\ref{fig:Fig2}(b).
It depicts roughly four periods on the horizontal time axis; the color encodes the voltage and the vertical axis indexes the lattice nodes (space), using the numbering convention of Fig.~\ref{fig:setup}(a).
It is evident that the profile is spatially localized over two unit cells, and we identify it as the CLS corresponding to this lowest flat band.
%PGK: what might have been more definitive here is also
%a *cut* showcasing the spatial profile at a given time
% vs. the theoretical eigenvector. This would be good
%to try for all 3 modes. Is this possible or not?

Similarly, Fig.~\ref{fig:Fig2}(c) depicts the spatial lattice response at the second flat band frequency of 154 kHz.
Again, the driver was injected at nodes 1 and 2.
This CLS mode is characterized by a vanishing amplitude at the center \(T\) site (site 3), with the \(\{U,V\}\) nodes to the left oscillating in-phase and the \(\{U,V\}\) nodes to the right oscillating in anti-phase.
Finally, Fig.~\ref{fig:Fig2}(d) shows the spatial signature at the highest flat band frequency of 244 kHz.
In contrast to the previous configuration, here the \(T\)-site (site 3) is highly excited, as expected from the previously discussed theoretical prediction of the CLS.
Furthermore, site 3 and sites 1, 2, and 4 are in anti-phase, which is expected from tight-binding calculation

We briefly comment on the quantitative quality of the data.
We use three metrics to assess data quality. 
(i) \(r = |T_n/\bar{X}_n|\), the absolute amplitude ratios of central site and peripheral sites, compared with the theory described in Sec.~\ref{experimental}.
Here, \(\bar{X}_n = \sum_{X_{n'} \in S-\{T_n\}}|X_{n'}|/4\) is the mean absolute amplitude, over the set \(S\) of the CLS sites (\(S =\{U_0, V_0, T_1, U_1, V_1\}\); 
and (ii)  \(\tilde\sigma = (\sigma_{S-\{T_1\}}/A\times 100) ~\%\), the absolute amplitude deviations at the perhipheral sites, which are expected to be zero in theory; 
and (iii) \(\textrm{SNR} = A/\sigma_{S^c} \), the signal-to-noise level outside the CLS, with \(S^c\) averaged over time and sites.
Here, \(A\) is the maximum amplitude of the CLS.

At 95 kHz (theory: 95 kHz), we obtain \(r = 1.23\) (theory: \(r = \sqrt 5 - 1 \approx 1.24\)), \(\tilde\sigma = 6~\%\), \(\textrm{SNR} \approx 15\).
At 154 kHz (theory: 152 kHz), we obtain \(r = 0.009\) (theory: 0), \(\tilde\sigma = 6~\%\), \(\textrm{SNR} \approx 100\).
At 244 kHz (theory: 244 kHz), we obtain \(r = 3.2\) (theory: \(r = \sqrt 5 + 1 \approx 3.2\)), \(\tilde\sigma = 5~\%\), \(\textrm{SNR} \approx 130\).
At 95 kHz the SNR is lower than for the other flat band resonances. 
Here \(T_n \approx -1.2 U_n\), which reduces the current of \(I_4\) and \(I_2\) at the input and output current of the negative inductive coupling, \(I_f = I_3 + I_4\) and \(I_i = I_1 + I_2\). 
The shorthand \(I_f \approx I_4\) and \(I_i \approx I_2\) remains valid for large \(R\), but is less accurate, and residual background becomes more visible.

\subsection{Excitations at the Edges}

Let us now cut the ring lattice to form a line with open boundary conditions.  
Note that the hopping terms that originally connected node~0 to nodes~13 and~14 should not be simply terminated;
they should be connected to ground to ensure the {correct} onsite potential at node~0, e.g. Eq.~\eqref{eq:ham-abf}. 
Similarly, the hopping terms that originally connected nodes~13 and~14 to node~0 should be connected to ground.

If we drive \(U_0\) and \(V_0\) (nodes~1 and~2) in-phase (the left edge), we excite a bulk CLS
away from the edge, so the edge has no effect. 
By contrast, if we drive nodes~1 and~2 in anti-phase, the CLS impinges on the edge. 
Tight-binding calculations predict two edge states at frequencies \(120\)~kHz and \(232\)~kHz.
These states are also compactly localized, occupying sites \(T_0\), \(U_0\), and \(V_0\) (nodes~0,~1, and~2), with \(U_0\) and \(V_0\) oscillating in anti-phase.

The blue shades of Figure~\ref{fig:Fig2}(e) illustrate the frequency response of anti-phase left edge driving. 
Two prominent peaks are observed at \(120\)~kHz and \(220\)~kHz, which agrees well with the tight-binding calculation. 
Panels (f)-(g) are the spatio-temporal steady-state profiles of the CLSs at those peaks, which show that they are compactly localized in space at (\(T_0\), \(U_0\), and \(V_0\)) (nodes~0,~1, and~2), as expected.
Furthermore, by comparing to Fig.~\ref{fig:Fig2}(b) and (c), we see that the patterns resemble truncated CLSs; the lower-frequency peak maps to the truncated acoustic CLS, whereas the higher-frequency peak to the truncated optic CLS.
The difference between them is simply whether the sites \(U_0\) and \(V_0\) (sites 0 and 1), which are connected via negative coupling, are oscillating in-phase or out-of-phase.

Next, we consider the right edge.  
Tight-binding calculations predict a single compactly localized edge mode localized on sites \(U_n\) and \(V_n\) at the frequency of the middle flat band, 152~kHz.  
Again, anti-phase driving produces a CLS that extends into the bulk, away from the edge, so the edge has no effect.
The red shade of figure \ref{fig:Fig2}(e) shows the frequency response at \(U_4\) (node~13) when two sites \(U_4\) and \(V_4\) (nodes~13 and~14) are driven in-phase.
When driving two nodes in-phase, we observe one prominent peak at \(150\)~kHz as predicted from the tight-binding calculation.
The spatial profile of the edge state is shown in Fig.~\ref{fig:Fig2}(h).
Only the nodes~13 and 14 are oscillating in phase, namely the ones that are driven.
Such a pattern is also accomplished via single-node driving at this frequency. If only site 13 or 14 is driven at that frequency, the same pattern is established.

We perform an analysis analogous to Sec.~\ref{sec:bulk}, with the only change that the peripheral set comprises two sites rather than four. 
In the tight-binding calculation, we enforce zero amplitude on the edge unit cells by imposing destructive-interference boundary conditions—\(\{U,V\}\) in phase on the left edge and anti-phase on the right—and then solve a \(3 \times 3\) eigenvalue problem.
At 120 kHz (theory: 120 kHz), we obtain \(r = 0.74\) (theory: \(r = \sqrt 3 - 1 \approx 0.73\)), \(\tilde\sigma = 0.5~\%\), \(\textrm{SNR} \approx 100\).
At 233 kHz (theory: 232 kHz), we obtain \(r = 2.89\) (theory: \(r = \sqrt 3 + 1 \approx 2.73\)), \(\tilde\sigma = 7~\%\), \(\textrm{SNR} \approx 62\).
At 150 kHz (theory: 152 kHz), we obtain \(r = 0.09\) (theory: \(r = 0\)), \(\tilde\sigma = 8~\%\), \(\textrm{SNR} \approx 150\).

\section{Conclusions and Future Challenges}

We have demonstrated the implementation of the \(\pi\)-flux diamond chain with all-bands-flat (ABF) using electric circuits, by devising negative inductive coupling that realizes negative hoppings.
The key properties of ABF were observed: complete absence of dispersion as evidenced by the lack of band curvature, 
\(k\)-independent resonant modes with compactly localized spatial profiles (CLS), or Aharonov-Bohm caging, as anticipated from the theory, in very good agreement between the two.
We also showed that not only can the electrical systems at hand showcase the relevant phenomenology of flat bands at the bulk, but also through suitably tailored predictions at the level of surface excitations, e.g.,
with homogeneous Dirichlet boundary conditions (i.e., experimental ``grounding'') at the left or/and right boundary.

Our experiment demonstrates the promising potential of electrical circuit networks as a simple, straightforward tunable platform to set and to spatio-temporally monitor modern band theory, including flat bands:
it can realize synthetic magnetic fluxes by inducing phase shifts (here, a \(\pi\) shift), demonstrating capability of realizing a complicated, fine-tuned tight-binding network such as an ABF one.
We believe that any flux, other than \(\pi\) can be also realized using op-amps, by implementing a suitable phase-shift instead of negating the voltage.

It will be interesting to explore the topological phases that might appear in such electrical networks.
For example, the diamond chain is also known to have symmetry-protected edge modes when the onsite potentials are all equal on all sites.
This can be easily realized by attaching additional inductors to the \(U_n, V_n\) nodes, and to the ground 
and may be an interesting extension for further studies.
Moreover, as already discussed previously, the present setting offers numerous additional opportunities for exploring the interplay of CLS-connected (compact) localization with impurity-induced Anderson-localization,
as has been recently explored also in photonic systems~\cite{dresselhaus2024tale}.
On the other hand, for sufficiently large voltage excitations, whereby elements such as the capacitors may operate in nonlinear capacitance-voltage dependent regimes~\cite{Remoissenet1999}, 
it is natural to explore the generalization of the configurations considered herein to the nonlinear regime.
The latter, as well as the nature of the associated excitations (compact or exponentially localized for different bands) and of the transport properties of the nonlinear lattice are particularly interesting directions also in their own right.

\bibliography{flat band,local}

\end{document}